\documentclass[10pt]{article}
\usepackage{latexsym}
\usepackage{amssymb}
\usepackage{amsmath}
\usepackage{amscd}
\usepackage{amsthm}
\usepackage[left=2cm,top=2.5cm,right=2.5cm,bottom=1.5cm]{geometry}

\usepackage[dvips]{graphicx}
\usepackage{hyperref}

\begin{document}
\begin{center}
\large{\bf {SOME LRS BIANCHI-I STRING COSMOLOGICAL MODELS WITH VARIABLE DECELERATION PARAMETER}} \\
\vspace{10mm}
\normalsize{Anil Kumar Yadav}\\
\vspace{4mm}
\normalsize{Department of Physics, Anand Engineering
College, Keetham, Agra - 282 007, India} \\
\vspace{2mm}
\normalsize{E-mail: abanilyadav@yahoo.co.in}\\
\end{center}
\vspace{10mm}
\begin{abstract} 
The present study deals with LRS Bianchi type I cosmological model
representing massive string. The energy-momentum tensor for such string as formulated by Letelier (1983) is used to construct 
massive string cosmological models for which we assume that the shear scalar ($\sigma$) is proportional 
to the expansion scalar ($\theta$). The study reveals that massive 
strings dominate the early Universe. The strings eventually disappear from the Universe for sufficiently 
large time, which is in agreement with the current astronomical observations. Some physical and geometrical 
behaviour of models are also discussed.\\       
\end{abstract}
\section{Introduction}
The past decades has been tremendous advances in cosmology.
The discovery of dark energy (Perlmutter et al \cite{ref1}$-$\cite{ref3}; Riess et al \cite{ref4}$-$\cite{ref5}) has crushed widely-held expectations 
that some unknown mechanism might set the cosmological constant to zero.
At the same time, substantial theoretical progress in string theory has brought
forth a diverse new generation of cosmological models, some of which are subject to
direct observational tests. One key advance in the emergence of methods of moduli stabilization. 
Compactification of string theory from the total dimension D down to four dimensions introduces many 
gravitationally-coupled scalar fields moduli from the point of view of the four dimensional theory. 
Recently we have studied inhomogeneous string cosmological model
formed by geometric string and use this model as a source of gravitational field \cite{ref6,ref7}. We had two main reason 
to study the above mentioned model. First, as a test of consistency, for some particular field theories based on string 
models and second we point out the Universe can be represented by a collection of extended galaxies. It is generally 
assumed that after the big bang, the Universe may have undergone a series of phase transitions as its temperature 
cooled below some critical 
temperature as predicted by grand unified theories \cite{ref8}$-$\cite{ref12}. At the very early stage of evolution 
of universe, it is believed that during the phase transition, the 
symmetry of Universe was broken spontaneously. That could have given rise to topologically-stable defects such as 
domain walls, strings and monopoles \cite{ref12}. Among all the three cosmological structures, only cosmic strings have excited 
the most interesting consequence \cite{ref13}, because it gives rise the density perturbations which leads 
to the formation of galaxies. The cosmic string can be closed (like loops) and open (like a hair) which move through
time and trace out a tube or a sheet, according to whether it is closed or open. These cosmic strings have stress 
energy and couple to the gravitational field. Therefore, it is interesting to study the gravitational effect 
which arises from strings by using Einstein's equations. \\ 

The general treatment of strings was initiated by Letelier \cite{ref14,ref15} and 
Stachel \cite{ref16}. Letelier \cite{ref14} obtained the general solution of Einstein's 
field equations for a cloud of strings with spherical, plane and a particular 
case of cylindrical symmetry. Letelier \cite{ref15} also obtained massive string 
cosmological models in Bianchi type-I and Kantowski-Sachs space-times. Benerjee 
et al \cite{ref17} have investigated an axially symmetric Bianchi type I string dust 
cosmological model in presence and absence of magnetic field using a supplementary 
condition $\alpha = a \beta$ between metric potential where $\alpha = \alpha(t)$ and 
$\beta = \beta(t)$ and $a$ is constant. Exact solutions of string cosmology for 
Bianchi type-II, $-VI_{0}$, -VIII and -IX space-times have been studied by Krori et al
\cite{ref18} and Wang \cite{ref19}. Wang \cite{ref20}$-$\cite{ref23} has investigated bulk viscous string 
cosmological models in different space-times. Bali and Anjali \cite{ref24}, Yadav \cite{ref25}, 
Pradhan et al \cite{ref26,ref27} and 
Yadav et al \cite{ref28} have studied string cosmological models in different physical contexts. 
The string cosmological models with a magnetic field are discussed by 
Chakraborty \cite{ref29}, Tikekar and Patel \cite{ref30,ref31}, Patel and Maharaj \cite{ref32}. Singh and Singh \cite{ref33} investigated string 
cosmological models with magnetic field in the context of space-time with $G_{3}$ 
symmetry. Singh \cite{ref34,ref35} has studied string cosmology with electromagnetic fields in 
Bianchi type-II, -VIII and -IX space-times. Lidsey, Wands and Copeland \cite{ref36} have 
reviewed aspects of super string cosmology with the emphasis on the cosmological 
implications of duality symmetries in the theory. Recently Saha and Visinescu \cite{ref37} and Saha et al \cite{ref38} 
have investigated Bianchi I string cosmological model in presence of magnetic flux. They have found that the present 
of cosmic string does not allow the anisotropic Universe to evolve into an isotropic one \cite{ref37}.\\

In this paper we have studied locally rotationally symmetric (LRS) Bianchi type I string cosmological model 
with time varying deceleration parameter (DP). The paper has following structure. In section 2, the metric and field equations are described. 
In section 3, we introduce a few plausible solutions consistent with observations. At the end we shall summarize the 
findings.\\ 

\section{The Metric and Field  Equations}
We consider the LRS Bianchi type I metric of the form
\begin{equation}
\label{eq1}
ds^2 = -dt^2 + A^2dx^2 + B^2 \left(dy^2 + dz^2\right)
\end{equation}
where, A and B are functions of t only. This ensures that the model is spatially homogeneous.\\

The energy-momentum tensor $T^{i}_{j}$ for a cloud
of massive strings and perfect fluid distribution is taken as
\begin{equation}
\label{eq2} T^{i}_{j} = (\rho + p)v^{i}v_{j} + p g^{i}_{j} -\lambda
x^{i}x_{j},
\end{equation}
where $p$ is the isotropic pressure; $\rho$ is the proper energy density for a cloud strings with particles
attached to them; $\lambda$ is the string tension density; $v^{i}=(0,0,0,1)$ is the four-velocity of the
particles, and $x^{i}$ is a unit space-like vector representing the direction of string. The vectors $v^{i}$
and $x^{i}$ satisfy the conditions
\begin{equation}
\label{eq3} v_{i}v^{i}=-x_{i}x^{i}=-1,\;\; v^{i}x_{i}=0.
\end{equation}

Choosing $x^{i}$ parallel to $\partial/\partial x$, we have
\begin{equation}
\label{eq4} x^{i} = (A^{-1},0,0,0).
\end{equation}
If the particle density of the configuration is denoted by
$\rho_{p}$, then
\begin{equation}
\label{eq5} \rho = \rho_{p}+\lambda.
\end{equation}
The Einstein's field equations (in gravitational units $c = 1$, $8\pi G = 1$) read as
\begin{equation}
\label{eq6}
R_{j}^i - \frac{1}{2}g_{j}^{i}R  = -T_{j}^i
\end{equation}
The Einstein's field equations (\ref{eq6}) for the line-element (\ref{eq1}) 
lead to the following system of equations
\begin{equation}
\label{eq7}
2\frac{B_{44}}{B} + \frac{B_{4}^2}{B^2}  = -p +\lambda
\end{equation}
\begin{equation}
\label{eq8}
\frac{A_{44}}{A} + \frac{B_{44}}{B} + \frac{A_{4}B_{4}}{AB} =-p  
\end{equation}
\begin{equation}
\label{eq9}
\frac{B_{4}^2}{B^2} + 2\frac{A_{4}B_{4}}{AB}  
 = \rho  
\end{equation}
Here, and in what follows, sub in-dices 4 in $A$, $B$ and elsewhere indicates
differentiation with respect to $t$. The energy conservation
equation $\;T^{ij}_{\;\;\;;j}=0$, leads to the following expression:
\begin{equation}
\label{eq10} \rho_{4} + (\rho + p)\left(\frac{A_{4}}{A} +
2\frac{B_{4}}{B}\right) -
\lambda\frac{A_{4}}{A} = 0\;,
\end{equation}
which is consequence of the field equations (\ref{eq7})-(\ref{eq9}).\\
The average scale factor (R) of LRS Bianchi type I model is defined as 
\begin{equation}
 \label{eq11}
R=(AB^{2})^{\frac{1}{3}}
\end{equation}
The spatial volume (V) is given by
\begin{equation}
\label{eq12}
V = R^{3} = AB^{2}
\end{equation}
We define the mean Hubble parameter (H) for LRS Bianchi I space-time as
\begin{equation}
\label{eq13}
H=\frac{R_{4}}{R}=\frac{1}{3}\left(\frac{A_{4}}{A}+2\frac{B_{4}}{B}\right)
\end{equation}
The expansion scalar ($\theta$), shear scalar ($\sigma$) and mean anisotropy parameter ($A_{m}$) are defined as
\begin{equation}
\label{eq14}
\theta =3H = \frac{A_{4}}{A}+2\frac{B_{4}}{B}\;, 
\end{equation}
\begin{equation}
\label{eq15} 
 \sigma^{2}=\frac{1}{2}\left(\sum_{i=1}^{3} H_{i}^{2}-\frac{1}{3}\theta^{2}\right)\;,
\end{equation}
\begin{equation}
\label{eq16}
A_{m} = \frac{1}{3}\sum_{i=1}^{3}\left(\frac{H_{i}-H}{H}\right)^{2}
\end{equation}

\section{Solutions of the Field Equations}
Observations of type Ia supernovae \cite{ref5} allow to probe the expansion history 
of Universe. In literature it is common to use a constant deceleration parameter, as it duly 
gives a power law for metric function or corresponding quantity. But at present the expansion of Universe 
is accelerating and decelerating in the past. Also the transition redshift from decelerating phase to accelerating 
phase is about $0.5$. Now for the Universe which was decelerating in the past and accelerating 
at present time, the DP must show signature flipping \cite{ref39}. So, in general, DP is not constant but variable. 
Following, Virey et al 
\cite{ref40}, we consider the DP to be variable i. e. 
\begin{equation}
 \label{eq17}
q=-\frac{RR_{44}}{R_{4}^{2}} = b\;(variable)
\end{equation}
where R is average scale factor. In this paper, we show, how the variable DP models with metric (\ref{eq1}) 
behave in presence of string fluid as a source of matter. \\

Pradhan et al \cite{ref6} and recently Yadav and Yadav \cite{ref41}, 
have obtained cosmological models with proportionality relation between 
shear scalar ($\sigma$) and expansion scalar ($\theta$). This condition
leads to the following relation between the metric potentials:  
\begin{equation}
\label{eq18} A = B^{n},
\end{equation}
where $n$ is a positive constant.\\
From equation (\ref{eq17}), we obtain
\begin{equation}
\label{eq19}
\frac{R_{44}}{R_{4}}+b\frac{R_{4}^{2}}{R^{2}} = 0
\end{equation}
In order to solve equation (\ref{eq19}), we have to assume $\;b = b\;(R)$. It is important to note here that one can assume 
$b = b\;(t) = b\;(R(t))$, as R is also a time dependent function. But this is possible only when one avoid singularity like big bang 
or big rip because both $t$ and $R$ are increasing function.\\
Thus the general solution of equation (\ref{eq19}) with assumption $\;b = b\;(R)$, is given by
\begin{equation}
\label{eq20}
\int{e^{\int\frac{b}{R}dR}} = t+m
\end{equation}
where m is the constant of integration.\\

One can not solve eq. (\ref{eq20}) in general as $b$ is variable. So, in order to solve the problem completely, 
we have to choose $\int\frac{b}{R}\;dR$ in such a manner that eq. (\ref{eq20}) be integrable with out any loss of generality, 
we consider
\begin{equation}
\label{eq21}
\int\frac{b}{R}dR = \emph{ln}\;L(R)
\end{equation}
Which does not effect the nature of generality of the solution.\\
Hence, from equations (\ref{eq20}) and (\ref{eq21}), we obtain
\begin{equation}
\label{eq22}
\int{L(R)dR}=t+m
\end{equation}
Of course, the choice of $L(R)$, in eq. (\ref{eq22}), is quite arbitrary but, since we are looking for a physically 
viable models of Universe consistent with observations. We consider the following cases 
\subsection{Solution in the polynomial form}
 Let us consider $L(R) = \frac{1}{2k_{1}\sqrt{R+k_{2}}}\;$, where $k_{1}$ and $k_{2}$ are constants.\\
In this case, on integrating, eq.(\ref{eq22}) gives the exact solution
\begin{equation}
\label{eq23}
R=\alpha_{1}T^{2} + \alpha_{2}T + \alpha_{3} 
\end{equation}
where\\
 $T=t+m$, $\alpha_{1} = k_{1}^{2}$, $\alpha_{2} = 2c_{1}k_{1}^{2}$, $\alpha_{3} = c_{1}^{2}k_{1}^{2}-k_{2}$\\ 
Here, $c_{1}$ is constant of integration.
Solving equations (\ref{eq11}), (\ref{eq18}) and (\ref{eq23}), we obtain the metric function as
\begin{equation}
\label{eq24}
A=\left(\alpha_{1}T^{2} + \alpha_{2}T + \alpha_{3}\right)^{\frac{3n}{n+2}}\;, 
\end{equation}
 \begin{equation}
\label{eq25}
B=\left(\alpha_{1}T^{2} + \alpha_{2}T + \alpha_{3}\right)^{\frac{3}{n+2}}\;, 
\end{equation} 
Hence the metric (\ref{eq1}) is reduced to
\begin{equation}
\label{eq26}
ds^2 = -dT^2 + \left(\alpha_{1}T^{2} + \alpha_{2}T + \alpha_{3}\right)^{\frac{6m}{m+2}}dx^2 + 
\left(\alpha_{1}T^{2} + \alpha_{2}T + \alpha_{3}\right)^{\frac{6}{m+2}} \left(dy^2 + dz^2\right)
\end{equation}

The expressions for the isotropic pressure ($p$), the proper energy
density ($\rho$), the string tension ($\lambda$) and the particle
density ($\rho_{p}$) for the model (\ref{eq26}) are obtained as
\begin{equation}
\label{eq27}
p=\frac{3(3n-2-5n^{2})(2\alpha_{1}T+\alpha_{2})^{2}}{(n+2)^{2}(\alpha_{1}T^{2} + \alpha_{2}T + \alpha_{3})^{2}}-
\frac{6\alpha_{1}(n+1)}{(n+2)(\alpha_{1}T^{2} + \alpha_{2}T + \alpha_{3})}\;,
\end{equation}
\begin{equation}
\label{eq28}
\rho = \frac{9(2n+1)(2\alpha_{1}T+\alpha_{2})^{2}}{(n+2)^{2}(\alpha_{1}T^{2} + \alpha_{2}T + \alpha_{3})^{2}}\;,
\end{equation}
\begin{equation}
\label{eq29}
\lambda = \frac{3(n+4-5n^{2})(2\alpha_{1}T+\alpha_{2})^{2}}{(n+2)^{2}(\alpha_{1}T^{2} + \alpha_{2}T + \alpha_{3})^{2}} -
\frac{6\alpha_{1}(n-1)}{(n+2)(\alpha_{1}T^{2} + \alpha_{2}T + \alpha_{3})}\;,
\end{equation}
\begin{equation}
\label{eq30}
\rho_{p} = \frac{3(5n^{2}+5n-1)(2\alpha_{1}T+\alpha_{2})^{2}}{(n+2)^{2}(\alpha_{1}T^{2} + \alpha_{2}T + \alpha_{3})^{2}} -
\frac{6\alpha_{1}(1-n)}{(n+2)(\alpha_{1}T^{2} + \alpha_{2}T + \alpha_{3})}\;,
\end{equation}
The energy conservation equation (\ref{eq10}) is satisfied identically by the above solutions, as expected.\\

We observe that all the parameters diverge at $T=\frac{-\alpha_{2}\pm\sqrt{\alpha_{2}^{2}-4\alpha_{1}\alpha_{3}}}{2\alpha_{1}}$. 
Therefore, the model has singularity 
at $T=\frac{-\alpha_{2}\pm\sqrt{\alpha_{2}^{2}-4\alpha_{1}\alpha_{3}}}{2\alpha_{1}}$, which can be shifted to $T=0$ by 
choosing $\alpha_{2} = \alpha_{3} = 0$. This singularity is of Point Type 
as all the scale factors vanish at $T=\frac{-\alpha_{2}\pm\sqrt{\alpha_{2}^{2}-4\alpha_{1}\alpha_{3}}}{2\alpha_{1}}$. 
The parameters $p$, $\rho$, $\rho_{p}$ and $\lambda$ start off 
with extremely large values. In particular, the large values of $\rho_{p}$ and $\lambda$ in the beginning suggest 
that strings dominate the early Universe. For sufficiently large time, $\rho_{p}$ and $\lambda$ become negligible. 
Therefore, the strings disappear from Universe for large time that is why, the strings are not observable in the present 
Universe.
From equation (\ref{eq28}), it is observed that the proper energy density $\rho$ is 
decreasing function of time. $\textbf{Fig. 1}$ depicts the variation of rest energy density versus time. 
The proper energy density $\rho$ and particle energy density $\rho_{p}$ have been graphed in $\textbf{Fig. 2}$. 
it is evident that $\rho_{p} > \rho$, i. e. the particle energy density remains larger than the proper energy density 
during the cosmic expansion, as expected.\\
 
From equation (\ref{eq29}) and (\ref{eq30}), we have $ \frac{\rho_{p}}{\arrowvert\lambda\arrowvert} > 1 $  i. e. particle 
energy density $(\rho_{p})$ remains larger than the string tension density $(\lambda)$ during the cosmic expansion, especially 
in early Universe. This behaviour of $\rho_{p}$ and $\arrowvert\lambda\arrowvert$ is clearly shown in $\textbf{Fig. 3}$. 
According to the ref. 
(Letelier \cite{ref15}; Krori et al \cite{ref18}),
 when $ \frac{\rho_{p}}{\arrowvert\lambda\arrowvert} > 1 $, in the process of evolution,
the Universe is dominated by massive strings, and when $ \frac{\rho_{p}}{\arrowvert\lambda\arrowvert} < 1 $, the 
universe is dominated by strings. From $\textbf{Fig. 3}$, we see that $ \frac{\rho_{p}}{\arrowvert\lambda\arrowvert} > 1 $. 
Thus in derived model, the early Universe is 
dominated by massive string. According to Ref. \cite{ref42}, since there is no direct evidence of strings in the present day Universe, we are in 
general, interested in constructing models of Universe that evolves purely from the era dominated by either geometric 
strings or massive strings and end up in the particle dominated era with or without remnants of strings. Therefore, 
the above model describes the evolution of the Universe consistent with the present-day observations.\\ 
\begin{figure}
\begin{center}
\includegraphics[width=4.0in]{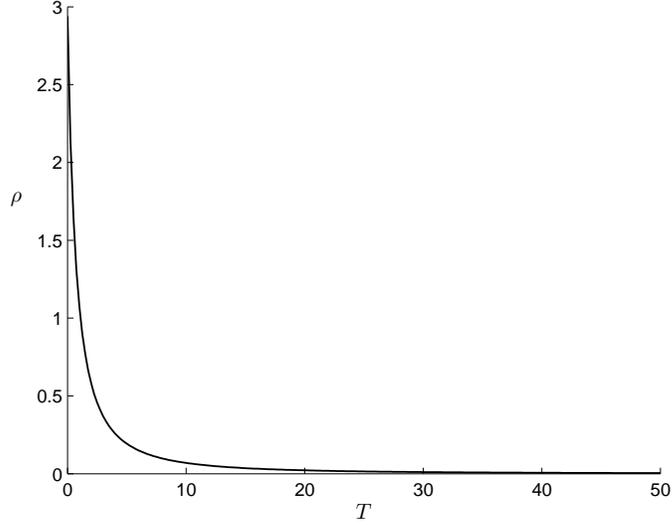} 
\caption{The plot of proper energy density $(\rho)$ vs. time (T).}
\label{fg:anil26RF2.eps}
\end{center}
\end{figure}
\begin{figure}
\begin{center}
\includegraphics[width=4.0in]{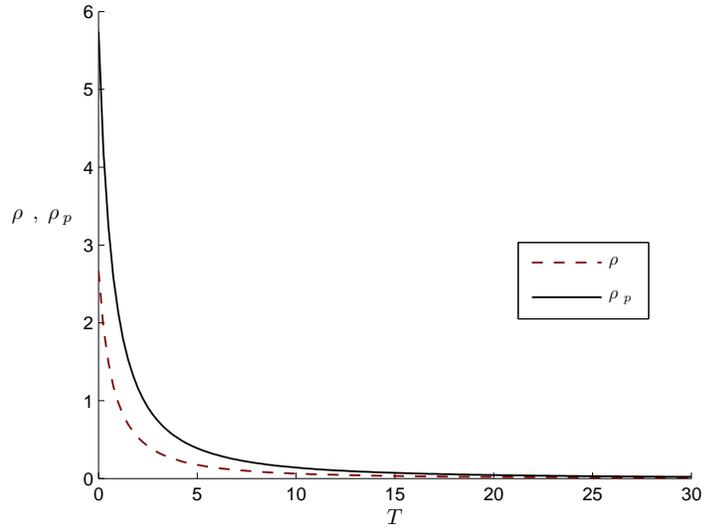} 
\caption{Proper energy density $(\rho)$ and particle energy density $(\rho_{p})$ vs. time (T).}
\label{fg:anil26RF3.eps}
\end{center}
\end{figure}
\begin{figure}
\begin{center}
\includegraphics[width=4.0in]{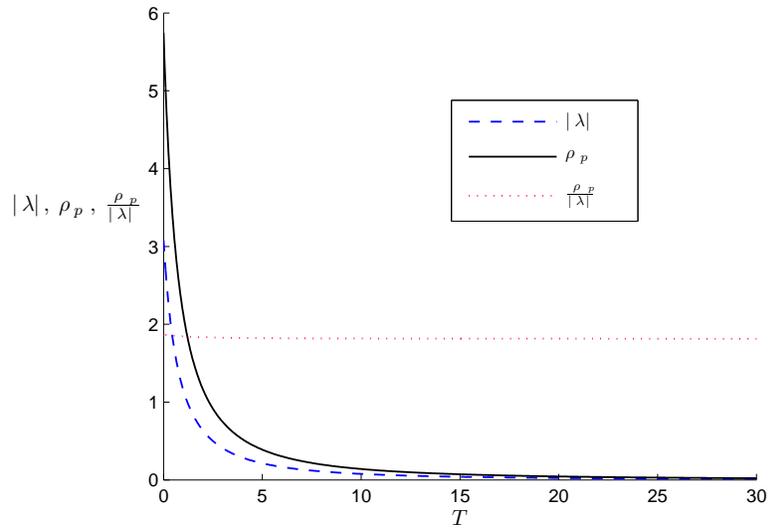} 
\caption{String tension density $(\lambda)$, particle energy density $(\rho_{p})$ and $ \frac{\rho_{p}}{\arrowvert\lambda\arrowvert}$ 
vs. time (T).}
\label{fg:anil26RF4.eps}
\end{center}
\end{figure}
The rate of expansion in the direction of $x$, $y$ and $z$ are given by
\begin{equation}
\label{eq31}
H_{x} = \frac{A_{4}}{A} = \frac{3n(2\alpha_{1}T+\alpha_{2})}{(n+2)(\alpha_{1}T^{2}+\alpha_{2}+\alpha_{3})} 
\end{equation}
\begin{equation}
\label{eq32}
H_{y} = H_{z} = \frac{3(2\alpha_{1}T+\alpha_{2})}{(n+2)(\alpha_{1}T^{2}+\alpha_{2}+\alpha_{3})} 
\end{equation}
The mean Hubble's parameter ($H$), expansion scalar $(\theta)$ and shear scalar $(\sigma^{2})$ of model (\ref{eq26}) 
are given by
\begin{equation}
\label{eq33}
H = \frac{(2\alpha_{1}T+\alpha_{2})}{(\alpha_{1}T^{2}+\alpha_{2}+\alpha_{3})} 
\end{equation}
\begin{equation}
\label{eq34}
\theta = \frac{3(2\alpha_{1}T+\alpha_{2})}{(\alpha_{1}T^{2}+\alpha_{2}+\alpha_{3})} 
\end{equation}
\begin{equation}
\label{eq35}
\sigma^{2} = \frac{3(n-1)^{2}(2\alpha_{1}T+\alpha_{2})^{2}}{(n+2)^{2}(\alpha_{1}T^{2}+\alpha_{2}+\alpha_{3})^{2}} 
\end{equation}
The spatial volume $(V)$, mean anisotropy parameter $(A_{m})$ and DP $(q)$ are found to be
\begin{equation}
\label{eq36}
V = (\alpha_{1}T^{2}+\alpha_{2}+\alpha_{3})^{3}
\end{equation}
\begin{equation}
\label{eq37}
A_{m} = \frac{2(n-1)^{2}}{(n+2)^{2}}
\end{equation}
\begin{equation}
\label{38}
q = -\frac{2\alpha_{1}(\alpha_{1}T^{2}+\alpha_{2}T+\alpha_{3})}{(2\alpha_{2}T+\alpha_{2})^{2}}
\end{equation}

The variation of DP versus time has been graphed in $\textbf{Fig. 4}$. It is observed that DP evolves with in the 
range predicted by present-day observations. From eq. (\ref{eq36}), it can be seen that the spatial volume is zero 
at $T=\frac{-\alpha_{2}\pm\sqrt{\alpha_{2}^{2}-4\alpha_{1}\alpha_{3}}}{2\alpha_{1}}$, and it increases with the cosmic time. 
The parameter $H_{x}$, $H_{y}$, $H_{z}$, $H$, $\theta$ and $\sigma^{2}$ diverge at initial singularity. These parameters 
decrease with evolution of Universe and finally drops to zero at late time. For $n=1$, the mean anisotropy parameter 
vanishes and the directional scale factors vary as\\
$$A = B = \left(\alpha_{1}T^{2} + \alpha_{2}T + \alpha_{3}\right)$$ 
Therefore, isotropy is achieved in the derived model for $n = 1$. For this particular values of $n$, we see that $A(T) = B(T) = R(T) $, 
therefore, metric (\ref{eq1}) reduces to flat FRW space-time. Thus, the derived model acquires flatness for $n = 1$. But in the 
same spirit, the string tension density $(\lambda)$ vanishes for $n=1$. Hence we can not choose $n=1$ in presence of string fluid 
as a source of matter in LRS Bianchi I space-time.  
\begin{figure}
\begin{center}
\includegraphics[width=4.0in]{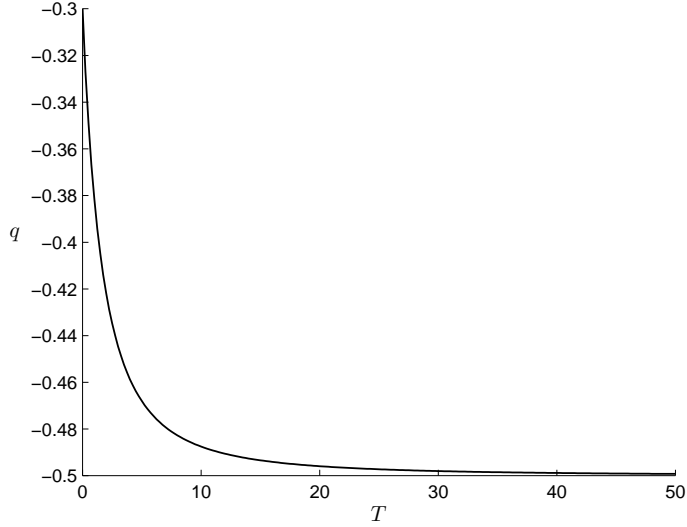} 
\caption{The plot of DP (q) vs. time (t).}
\label{fg:anil26RF1.eps}
\end{center}
\end{figure}

\subsection{Solution in sine hyperbolic form} 
We consider, $L(R) = \frac{1}{k_{3}\sqrt{1+R^{2}}}\;$, where $k_{3}$ is an arbitrary constant.\\
In this case, on integrating, eq.(\ref{eq22}) gives the exact solution
\begin{equation}
\label{eq39}
R=sinh\;(k_{3}T)
\end{equation}
where $T=t+m$, and the constant of integration has been omitted by assuming that $R=0$ at $T=0$.\\
Solving equations (\ref{eq11}), (\ref{eq18}) and (\ref{eq39}), we obtain the metric function as
\begin{equation}
\label{eq40}
A=sinh^{\frac{3n}{n+2}}\;(k_{3}T)
\end{equation}
\begin{equation}
\label{41}
B=sinh^{\frac{3}{n+2}}\;(k_{3}T)
\end{equation}
Hence the metric (\ref{eq1}) is reduced to
\begin{equation}
\label{eq42}
ds^2 = -dT^2 + sinh^{\frac{6n}{n+2}}(\;k_{3}T)dx^2 + sinh^{\frac{6}{n+2}}(\;k_{3}T) \left(dy^2 + dz^2\right)
\end{equation}

The expressions for the isotropic pressure ($p$), the proper energy
density ($\rho$), the string tension ($\lambda$) and the particle
density ($\rho_{p}$) for the model (\ref{eq42}) are obtained as
\begin{equation}
\label{eq43}
p=\frac{3(n+1)k_{3}^{2}}{n+2}cosech^{2}(\;k_{3}T)-\frac{9(n^{2}+n+1)k_{3}^{2}}{(n+2)^{2}}coth^{2}(\;k_{3}T)\;,
\end{equation}
\begin{equation}
\label{eq44}
\rho=\frac{9(2n+1)k_{3}^{2}}{(n+2)^{2}}coth^{2}(\;k_{3}T)\;,
\end{equation}
\begin{equation}
\label{eq45}
\lambda = \frac{3(n-1)k_{3}^{2}}{n+2}cosech^{2}(\;k_{3}T)-\frac{9(n^{2}+n-2)k_{3}^{2}}{(n+2)^{2}}coth^{2}(\;k_{3}T)\;,
\end{equation}
\begin{equation}
\label{eq46}
\rho_{p}=\frac{9(n^{2}+3n-1)k_{3}^{2}}{(n+2)^{2}}coth^{2}\;(k_{3}T)-\frac{3(n-1)k_{3}^{2}}{n+2}cosech^{2}(\;k_{3}T)\;,
\end{equation}
The energy conservation equation (\ref{eq10}) is satisfied identically by the above solutions, as expected.\\

We observe that all the parameters diverge at $T=0$. 
Therefore, the model has big bang singularity 
at $T=0$. This singularity is of Point Type 
as all the scale factors vanish at $T=0$. 
The parameters $p$, $\rho$, $\rho_{p}$ and $\lambda$ start off 
with extremely large values.
\begin{figure}
\begin{center}
\includegraphics[width=4.0in]{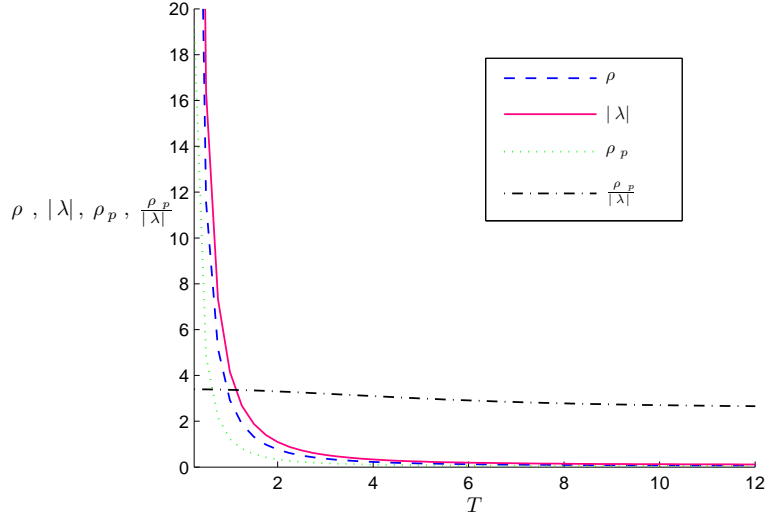} 
\caption{Proper energy density $(\rho)$, string tension density $(\lambda)$, particle energy density $(\rho_{p})$ 
and $ \frac{\rho_{p}}{\arrowvert\lambda\arrowvert}$ vs. time (T).}
\label{fg:anil26RF6.eps}
\end{center}
\end{figure}

The proper energy density $\rho$, string tension density $\lambda$ and particle energy density $\rho_{p}$ have 
been graphed versus time $T$ in $\textbf{Fig. 5}$. It is evident that string tension density becomes negligible 
for sufficient large time. Therefore, the strings disappear from Universe at late time that is why, the strings are not 
observable in present universe. From $\textbf{Fig. 5}$, we see that $ \frac{\rho_{p}}{\arrowvert\lambda\arrowvert} > 1 $. 
Therefore, the early Universe was 
dominated by massive string.\\
\begin{figure}
\begin{center}
\includegraphics[width=4.0in]{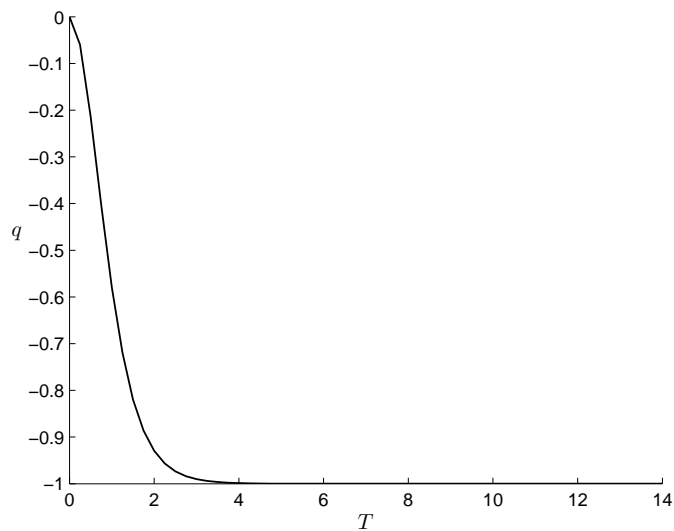} 
\caption{The plot of DP (q) vs. time (T).}
\label{fg:anil26RF5.eps}
\end{center}
\end{figure}
The rate of expansion in the direction of $x$, $y$ and $z$ are given by
\begin{equation}
\label{eq47}
H_{x} = \frac{3nk_{3}}{n+2}coth\;(k_{3}T)\;,
\end{equation}
\begin{equation}
\label{eq48}
H_{y}=H_{z}= \frac{3k_{3}}{n+2}coth\;(k_{3}T)\;,
\end{equation}
The mean Hubble's parameter ($H$), expansion scalar $(\theta)$ and shear scalar $(\sigma^{2})$ of model (\ref{eq42}) 
are given by
\begin{equation}
\label{eq49}
H=k_{3}coth\;(k_{3}T)\;,
\end{equation}
\begin{equation}
\label{eq50}
\theta=3k_{3}coth\;(k_{3}T)\;,
\end{equation}
\begin{equation}
\label{eq51}
\sigma^{2}=\frac{3(n-1)^{2}k_{3}^{2}}{(n+2)^{2}}coth^{2}\;(k_{3}T)
\end{equation}
The spatial volume $(V)$, mean anisotropy parameter $(A_{m})$ and DP $(q)$ are found to be
\begin{equation}
\label{eq52}
V=sinh^{3}\;(k_{3}T)\;,
\end{equation}
\begin{equation}
\label{eq53}
A_{m}=\frac{2(n-1)^{2}}{(n+2)^{2}}\;,
\end{equation}
\begin{equation}
\label{eq54} 
q=-tanh^{2}\;(k_{3}T)
\end{equation}
The variation of DP versus time has been graphed in $\textbf{Fig. 6}$. It is observed that DP evolves with in the range 
predicted by SN Ia \cite{ref1}$-$\cite{ref5} and CMBR \cite{ref43} observations. We observe that at $T=0$, the spatial 
volume vanishes and it increases with cosmic time. For $n=1$, the mean anisotropy parameter vanishes and the directional 
scale factors vary as 
$$A(T) = B(T) = sinh\;(k_{3}T)$$
Therefore, $n=1$, turns out to condition of isotropy but in the same spirit, the string tension density $(\lambda)$ vanishes. 
Therefore, the presence of cosmic string does not allow to choose $n = 1$.
\section{Concluding Remarks}
In this paper, we have studied LRS Bianchi type I string cosmological models in general relativity. 
The Einstein's field equations have been solved exactly with suitable physical assumptions and the solutions satisfy 
the energy conservation equation identically. Therefore, exact and physically viable LRS Bianchi I string cosmological models 
have been obtained. The derived models have singular origin i. e. the Universe starts expanding with a big bang singularity.\\

The main features of the models are as follows:\\
\begin{itemize}
 \item The models are based on exact solution of Einstein's field equations for LRS Bianchi I space-time 
in presence of string fluid as a source of matter.\\

\item It has been found that massive strings dominate the early Universe, which is eventually disappear from 
the Universe for sufficiently large time. This is in agreement with current astronomical observations.\\

\item In the derived models, $n=1$, turns out to be a condition of isotropy and flatness of Universe. 
It is important to mention here that for $n = 1$, the string tension density $(\lambda)$ vanishes in both cases. 
So, we conclude that presence of cosmic string does not allow to choose $n = 1$. Thus, the 
anisotropic LRS Bianchi I Universe may not evolve into isotropic one in presence of cosmic string. 
The same is predicted by Saha and Visinescu \cite{ref37} 
with different approach in Bianchi I space-time.     \\

\item The DP $(q)$ is evolving with negative value and the existing range of $q$ is in nice agreement with SN Ia 
data and CMBR observations. Thus the derived models are realistic.

\end{itemize}

\section*{Acknowledgements} 
Author is thankful to The Institute of Mathematical Science (IMSc), Chennai, India 
for providing facility and support where part of this work was carried out. Also Author thanks to B. Saha for helpful 
discussions.



\begin{thebibliography}{000}
\bibitem{ref1}
Permutter, S., \emph{et al}: Astrophys. J. {\bf 483}, 565 (1997)  
\bibitem{ref2}
Permutter, S., \emph{et al}: Nature {\bf 391}, 51 (1998)  
\bibitem{ref3}
Permutter, S., \emph{et al}: Astrophys. J. {\bf 517}, 565 (1999)   
\bibitem{ref4}
Ries, A. G., \emph{et al}: Astron. J. {\bf 116}, 1009 (1998) 
\bibitem{ref5}
Ries, A. G., \emph{et al}: Astron. J. {\bf 607}, 665 (2004)
\bibitem {ref6}
Pradhan, A., Yadav, A. K., Singh, R .P., and Singh, V. K.: Astrophys Space Sci. {\bf 312}, 145 (2007) 
\bibitem {ref7}
Yadav, A. K., Yadav, V. K., and Yadav, L.: Int. J. Theor. Phys. {\bf 48}, 568 (2009)  
\bibitem {ref8}
Zel'dovich, Ya. B., Kobzarev, I. Yu. and Okun, L. B.: Zh. Eksp. Teor. Fiz. 
{\bf 67}, 3 (1975)  
\bibitem {ref9}
Kibble, T. W. B.: J. Phys. A: Math. Gen. {\bf 9}, 1387 (1976)
\bibitem {ref10}  
Kibble, T. W. B.: Phys. Rep. {\bf 67},  183 (1980) 
\bibitem {ref11}
Everett, A. E.: Phys. Rev. {\bf 24}, 858 (1981) 
\bibitem {ref12}
Vilenkin, A.: Phys. Rev. D {\bf 24}, 2082 (1981) 
\bibitem {ref13}
Vilenkin, A.: Phys. Rep. {\bf 121}, 263 (1985)   
\bibitem {ref14}
Letelier, P. S.: Phys. Rev. D {\bf 20}, 1249 (1979)  
\bibitem {ref15}
Letelier, P. S.: Phys. Rev. D {\bf 28}, 2414 (1983)  
\bibitem {ref16}
Stachel,J.: Phys. Rev. D {\bf 21}, 2171 (1980) 
\bibitem {ref17}
Banerjee, A., Sanyal, A. K., and Chakraborty, S.: Pramana-J. Phys. {\bf 34}, 1 (1990)  
\bibitem {ref18}
Krori, K. D., Chaudhury,T., Mahanta, C. R. and Mazumder, A.:Gen. Rel. Grav. {\bf 22}, 123 (1990) 
\bibitem {ref19}
Wang, X. X.: Chin. Phys. Lett. {\bf 20}, 615 (2003)  
\bibitem {ref20}
Wang, X. X.: Astrophys. Space Sci. {\bf 293}, 933 (2004)  
\bibitem {ref21}
Wang, X. X.: Chin. Phys. Lett. {\bf 21}, 1205 (2004)  
\bibitem {ref22}
Wang, X. X.: Chin. Phys. Lett. {\bf 22}, 29 (2005) 
\bibitem {ref23}
Wang, X. X.: Chin. Phys. Lett. {\bf 23}, 1702 (2006)   
\bibitem {ref24}
Bali, R. and Anjali: Astrophys. Space Sci. {\bf 302}, 201 (2006) 
\bibitem {ref25}
Yadav, A. K.: arXiv: 0912.4801 [gr-qc] (2009)
\bibitem {ref26}
Pradhan, A., Mishra, M. K. and Yadav, A. K.: Rom. J. Phys. {\bf 54}, 747 (2009) 
\bibitem {ref27}
Pradhan, A., Agarwal, S. and Yadav, A. K.: Rom. J. Phys. {\bf 54}, 15 (2009) 
\bibitem {ref28}
Yadav, A. K., Yadav, V. K. and Yadav, L.: \emph{Pramana-J. Phys.} {\bf 76}, 681 (2011)
\bibitem {ref29}
Chakraborty, S.: Ind. J. Pure Appl. Phys. {\bf 29}, 31 (1980)  
\bibitem {ref30}
R. Tikekar and L. K. Patel, \emph{Gen. Rel. Grav.} {\bf 24}, 397 (1992) 
\bibitem {ref31} 
Tikekar, R. and Patel, L. K.: Pramana-J. Phys. {\bf 42}, 483 (1994) 
\bibitem {ref32}
Patel, L. K. and Maharaj, S. D.: Pramana-J. Phys. {\bf 47}, 1 (1996)
\bibitem {ref33}
Singh, G. P. and Singh, T.: Gen. Rel. Grav. {\bf 31}, 371 (1999)  
\bibitem {ref34}
Singh, G. P.: Nuovo Cim. B {\bf 110}, 1463 (1995) 
\bibitem {ref35}
Singh, G. P.: Pramana-J. Phys. {\bf 45}, 189 (1995) 
\bibitem {ref36}
Lidsey, J. E., Wands, D. and Copel, E. J.: Phys. Rep. {\bf 337}, 343 (2000)  
\bibitem{ref37}
Saha, B. and Visinescu, M.: Int. J. Theor. Phys. {\bf 49}, 1411 (2010)  
\bibitem{ref38}
Saha, B., Rikhvitsky, V. and Visinescu, M.: Cent. Eur. J. Phys. {\bf 8}, 113 (2010) 
\bibitem{ref39}
Padmanabhan, T. and Roychowdhury, T.: \emph{Mon. Not. R. Astron. Soc.} {\bf 344}, 823 (2003)
\bibitem{ref40}
Virey, J.M., \emph{et al}: Phys. Rev. D {\bf 72}, 061302 (2005) 
\bibitem{ref41}
Yadav, A. K. and Yadav, L.: Int. J. Theor. Phys. {\bf 50}, 218 (2011)
\bibitem{ref42}
Belinchon,  J. A.: Astrophys. Space Sci {\bf 323}, 185 (2009) 
\bibitem{ref43}
Bennett, C. L.,\emph{et al}: Astrophys. J. Suppl. {\bf 148}, 1 (2003)


\end{thebibliography}
\end{document}